\documentclass[hidelinks]{ieeetj}
\usepackage{cite}
\usepackage{amsmath,amssymb,amsfonts}
\usepackage{graphicx,color}
\usepackage{textcomp}
\usepackage{xcolor}
\usepackage{hyperref}
\usepackage{algorithm,algpseudocode}

\usepackage[acronym,shortcuts]{glossaries}
\usepackage{bm}
\usepackage{lipsum}
\usepackage{mleftright}
\usepackage{orcidlink}
\usepackage{booktabs}
\usepackage{mathtools}
\usepackage{fix-cm}

\def\BibTeX{{\rm B\kern-.05em{\sc i\kern-.025em b}\kern-.08em
    T\kern-.1667em\lower.7ex\hbox{E}\kern-.125emX}}
\AtBeginDocument{\definecolor{tmlcncolor}{cmyk}{0.93,0.59,0.15,0.02}\definecolor{NavyBlue}{RGB}{0,86,125}}

\def\OJlogo{\vspace{-4pt}$<$Society logo(s) and publication title will appear here.$>$}
\def\seclogo{\vspace{10pt}$<$Society logo(s) and publication title will appear here.$>$}

\def\authorrefmark#1{\ensuremath{^{\textbf{#1}}}}

\newacronym{SVD}{SVD}{singular value decomposition}
\newacronym{DCM}{DCM}{double-centering matrix}
\newacronym{3D}{3D}{three-dimensional}
\newacronym{GA}{GA}{genie-aided}
\newacronym{EA}{EA}{``\emph{estimate-then-average}''}
\newacronym{AE}{AE}{``\emph{average-then-estimate}''}
\newacronym{IRS}{IRS}{intelligent reflecting surface}
\newacronym{RSSI}{RSSI}{received signal strength indicator}
\newacronym{SotA}{SotA}{state-of-the-art}
\newacronym{CSI}{CSI}{channel state information}
\newacronym{D2D}{D2D}{device-to-device}
\newacronym{RR}{RR}{round-robin}
\newacronym{DA}{DA}{Dutch auction}
\newacronym{AV}{AV}{autonomous vehicle}
\newacronym{CWFL}{CWFL}{clustered WFL}
\newacronym{WFL}{WFL}{wireless federated learning}
\newacronym{RSMA}{RSMA}{rate splitting multiple access}
\newacronym{IoT}{IoT}{Internet-of-Things}
\newacronym{TDMA}{TDMA}{time-domain multiple access}
\newacronym{NOMA}{NOMA}{non-orthogonal multiple access}
\newacronym{ML}{ML}{machine learning}
\newacronym{MIMO}{MIMO}{multiple-input multiple-output}
\newacronym{CT}{CT}{compute-then-transmit}
\newacronym{FP}{FP}{fractional programming}
\newacronym{CF-mMIMO}{CF-mMIMO}{cell free massive MIMO}
\newacronym{iid}{i.i.d.}{independent and identically distributed}
\newacronym{AD}{AD}{autonomous driving}
\newacronym{DL}{DL}{downlink}
\newacronym{UL}{UL}{uplink}
\newacronym{IC}{IC}{interference cancellation}
\newacronym{SIC}{SIC}{successive interference cancellation}
\newacronym{BS}{BS}{base station}
\newacronym{TX}{TX}{transmit}
\newacronym{RX}{RX}{receive}
\newacronym{MU}{MU}{multi-user}
\newacronym{SISO}{SISO}{single-input single-output}
\newacronym{AWGN}{AWGN}{additive white Gaussian noise}
\newacronym{SINR}{SINR}{signal-to-interference-and-noise ratio}
\newacronym{FL}{FL}{federated learning}
\newacronym{CPU}{CPU}{central processing unit}
\newacronym{KNN}{KNN}{K-nearest-neighbor}
\newacronym{RF}{RF}{radio frequency}
\newacronym{GD}{GD}{gradient descent}
\newacronym{V2X}{V2X}{vehicle-to-anything}
\newacronym{i.i.d.}{i.i.d.}{independent and identically distributed}

\newacronym{RSS}{RSS}{received signal strength}
\newacronym{FIM}{FIM}{fisher information matrix}
\newacronym{ToA}{ToA}{time of arrival}
\newacronym{ToF}{ToF}{time of flightl}
\newacronym{AoA}{AoA}{angle of arrival}
\newacronym{GP}{GP}{Gaussian process}
\newacronym{2D}{2D}{two-dimensional}
\newacronym{GPR}{GPR}{Gaussian process regression}
\newacronym{GNSS}{GNSS}{global navigation satellite systems}
\newacronym{B5G}{B5G}{beyond fifth-generation}
\newacronym{6G}{6G}{sixth-generation}

\newacronym{RRH}{RRH}{remote radio head}
\newacronym{GPS}{GPS}{Global Positioning System}
\newacronym{RFID}{RFID}{radio frequency identification}
\newacronym{TCAS}{TCAS}{traffic alert and collision avoidance systems}
\newacronym{RMSE}{RMSE}{root mean square error}
\newacronym{MSE}{MSE}{mean square error}

\newacronym{SGD}{SGD}{stochastic gradient descent}
\newacronym{PDF}{PDF}{probability density function}
\newacronym{CU}{CU}{computing unit}
\newacronym{DM-MIMO}{DM-MIMO}{distributed massive multiple-input multiple-output}
\newacronym{LOS}{LOS}{line-of-sight}
\newacronym{NLOS}{NLOS}{non-line-of-sight}
\newacronym{ROI}{ROI}{region of interest}
\newacronym{AP}{AP}{access point}
\newacronym{TDOA}{TDOA}{time difference of arrival}
\newacronym{UE}{UE}{user equipment}
\newacronym{dB}{dB}{decibel}
\newacronym{RIS}{RIS}{reconfigurable intelligent surface}
\newacronym{CG}{CG}{conjugate gradient}

\newacronym{PG}{PG}{proximal gradient}
\newacronym{SVT}{SVT}{singular value thresholding}
\newacronym{NN}{NN}{nuclear norm}
\newacronym{NMSE}{NMSE}{normalized mean square error}
\newacronym{MC}{MC}{matrix completion}
\newacronym{NP}{NP}{non-deterministic polynomial-time}
\newacronym{EDM}{EDM}{euclidean distance matrix}
\newacronym{SC}{SC}{soft-connected}
\newacronym{CRLB}{CRLB}{Cramér-Rao Lower Bound}
\newacronym{PoA}{PoA}{phase of arrival}
\newacronym{UAV}{UAV}{unmanned aerial vehicle}
\newacronym{VR}{VR}{virtual reality}
\newacronym{MDS}{MDS}{multidimensional scaling}
\newacronym{SMDS}{SMDS}{super multidimensional scaling}

\newacronym{RBL}{RBL}{rigid body localization}
\newacronym{RBT}{RBT}{rigid body tracking}
\newacronym{SC-RBL}{SC-RBL}{soft-connected RBL}
\newacronym{W-RBL}{W-RBL}{\underline{wireless} RBL}

\newacronym{SDP}{SDP}{semidefinite programming}
\newacronym{JCAS}{JCAS}{joint communication and sensing}
\newacronym{SDR}{SDR}{semi-definite relaxation}

\newacronym{OPP}{OPP}{orthogonal Procrustes problem}
\newacronym{SLAM}{SLAM}{simultaneous localization and mapping}
\newacronym{WLS}{WLS}{weighted least square}
\newacronym{SI}{SI}{soft-impute}
\newacronym{GaBP}{GaBP}{Gaussian belief propagation}

\newacronym{SGA}{SGA}{scalar Gaussian approximation}

\newacronym{6D}{6D}{sixth-dimensional}
\newacronym{FDOA}{FDOA}{frequency difference of arrival}

\newacronym{wlg}{w.l.g.}{without loss of generality}
\newacronym{sIC}{soft-IC}{soft interference cancellation}
\newacronym{WL}{WL}{wireless localization}
\newacronym{XR}{XR}{extended reality}
\newacronym{IMU}{IMU}{inertial measurement unit}

\newacronym{LS}{LS}{least squares}

\newacronym{TC}{TC}{tensor completion}

\DeclareMathOperator*{\argmin}{arg\,min}

\begin{document}
\receiveddate{XX Month, XXXX}
\reviseddate{XX Month, XXXX}
\accepteddate{XX Month, XXXX}
\publisheddate{XX Month, XXXX}
\currentdate{XX Month, XXXX}
\doiinfo{XXXX.2022.1234567}

\markboth{}{Niclas F\"uhrling {et al.}}

\title{Discrete Aware Tensor Completion \\via Convexized $\ell_0$-Norm Approximation
\vspace{-.5ex}}

\author{Niclas~F\"uhrling\authorrefmark{1}\textsuperscript{\orcidlink{0000-0003-1942-8691}}, Graduate Student Member, IEEE,  Getuar Rexhepi\authorrefmark{1}\textsuperscript{\orcidlink{0009-0002-3268-522X}}, \\Graduate Student Member, IEEE, Giuseppe Abreu\authorrefmark{1}\textsuperscript{\orcidlink{0000-0002-5018-8174}}, Senior Member, IEEE}
%, David~Gonz{\'a}lez~G.\authorrefmark{2}\textsuperscript{\orcidlink{0000-0003-2090-8481}}, Senior Member, IEEE, and Osvaldo~Gonsa\authorrefmark{2}\textsuperscript{\orcidlink{0000-0001-5452-8159}}}
\affil{School of Computer Science and Engineering, Constructor University, Bremen, Germany}
%\affil{Wireless Communications Technologies Group, Continental AG, Frankfurt, Germany}
\corresp{Corresponding author: Niclas F\"uhrling (email: nfuehrling@constructor.university).}

\begin{abstract}
We consider a novel algorithm, for the completion of partially observed low-rank tensors, where each entry of the tensor can be chosen from a discrete finite alphabet set, such as in common image processing problems, where the entries represent the RGB values.
The proposed low-rank \ac{TC} method builds on the conventional \ac{NN} minimization-based low-rank \ac{TC} paradigm, through the addition of a discrete-aware regularizer, which enforces discreteness in the objective of the problem, by an $\ell_0$-norm regularizer that is approximated by a continuous and differentiable function normalized via \ac{FP} under a \ac{PG} framework, in order to solve the proposed problem.
Simulation results demonstrate the superior performance of the new method both in terms of \ac{NMSE} and convergence, compared to the conventional \ac{SotA} techniques, including \ac{NN} minimization approaches, as well as a mixture of the latter with a matrix factorization approach.

\end{abstract}


\maketitle

\glsresetall

\section{Introduction}

\IEEEPARstart{W}{ith} the recent rise of machine learning and big data, \ac{MC} has emerged as a fundamental problem in a wide range of modern applications, including recommendation systems in computer science \cite{Chen_2022}, localization algorithms in signal processing \cite{Nguyen_2019_Loc}, and millimeter-wave channel estimation in wireless communications \cite{Vlachos_2018}.

Most practical instances can be formulated as structured low-rank \ac{MC} problems, where the goal is to recover a low-rank matrix $\boldsymbol{X}\in\mathbb{R}^{m\times n}$ from a partially observed and incomplete matrix $\boldsymbol{O}\in\mathbb{R}^{m\times n}$ \cite{Nguyen_2019 , Dai_2012 , Bart_2013}. Classical \ac{MC} approaches \cite{Candes_2009_Noise,Candes_2009,Candes_2010} address this problem by replacing the inherently non-convex rank minimization objective with a convex surrogate, namely the \ac{NN}, in order to improve tractability and enable theoretical recovery guarantees.

Several \ac{SotA} methods follow this paradigm. In particular, the approach proposed in \cite{Candes_2009_Noise} employs the \ac{NN} as a tight convex relaxation of the rank function and demonstrates reliable recovery even from a small number of noisy observations.
Similarly, \cite{Candes_2009} considers \ac{NN} minimization solved via \ac{SDP}.
However, the computational complexity of \ac{SDP}-based solutions scales as $\mathcal{O}(\text{max}(m,n)^4)$, which severely limits their applicability to large-scale problems.

To overcome this computational burden, the method introduced in \cite{Cai_2010} leverages \ac{SVT} within a \ac{PG} framework to efficiently minimize the \ac{NN} objective, significantly reducing complexity while maintaining strong empirical performance.

Motivated by these developments, we recently proposed in \cite{Iimori_2020} and \cite{Nic_Asilo_2024} a related \ac{PG}-based, low-complexity low-rank \ac{MC} framework addressing a more challenging setting in which the matrix entries are constrained to belong to a finite discrete alphabet.
The resulting discrete-aware \ac{MC} approach is particularly well suited for applications such as recommendation and rating systems \cite{Chen_2022}.
Discreteness is incorporated by augmenting the optimization problem with a discrete-space regularizer, yielding closed-form updates under relaxed formulations based on an $\ell_1$-norm convexification \cite{Iimori_2020}, as well as a smooth non-convex approximation of the $\ell_0$-norm \cite{Nic_Asilo_2024}.

More recently, however, the increasing prevalence of multiway data has motivated the extension of matrix completion to higher-order structures, leading to the problem of tensor completion (TC). In many modern applications, observations are inherently multidimensional and cannot be faithfully represented in matrix form without losing structural information. 
Examples include signal processing, machine learning and data analysis, as well as scientific computing \cite{Song_2019,Ji_2019,Wang_2025}, where measurements naturally form tensors of order three or higher. 
In this context, matrix completion can be viewed as a special case of tensor completion restricted to second-order tensors.

Tensor completion aims to recover a low-rank tensor $\mathcal{\boldsymbol{T}} \in \mathbb{R}^{I_1 \times I_2 \times \cdots \times I_N},$ from a partially observed tensor $\mathcal{\boldsymbol{O}}$ of the same dimensions, under the assumption that the true tensor admits a low-dimensional latent structure. Unlike matrices, however, tensors admit multiple notions of rank, such as CP rank, Tucker rank, and tubal rank, each capturing different types of multilinear dependencies. 
This fundamental difference makes TC both more expressive and significantly more challenging than its matrix counterpart.

Early TC methods largely mirror developments in MC by replacing non-convex tensor rank objectives with convex or tractable surrogates. 
A common approach relies on minimizing sums of nuclear norms of mode-n unfoldings, effectively extending NN-based MC techniques to the tensor setting, as proposed in \cite{Liu_2013}. 
While such relaxations preserve convexity and theoretical guarantees, they often suffer from high computational complexity and may fail to fully exploit the intrinsic multilinear structure of the tensor. 
As an alternative, factorization-based formulations, which parameterize the tensor through low-dimensional latent factors, have been widely adopted due to their favorable scalability and reduced memory footprint \cite{xu2013parallel}. 
These approaches generalize matrix factorization techniques to higher-order tensors and typically lead to non-convex optimization problems that are nevertheless efficiently solvable in practice via iterative algorithms.

In addition, hybrid methods combining nuclear-norm regularization with factorized representations have been proposed to bridge the gap between convex relaxations and scalable low-rank modeling \cite{Gao2018}. 
Such formulations retain some of the regularization benefits of \ac{NN}-based approaches while significantly lowering computational complexity, thereby offering a flexible trade-off between theoretical tractability and practical performance.

Overall, tensor completion provides a principled and powerful generalization of matrix completion, enabling the recovery of incomplete data in scenarios where multidimensional correlations play a central role. 
This added modeling flexibility comes at the cost of increased algorithmic and theoretical complexity, motivating ongoing research into low-complexity, structure-aware TC methods that extend the successes of modern MC techniques to higher-order settings.

In light of the above, in this article, we propose an extension of the \ac{SotA} \acf{TC} techniques via a novel discrete-aware low-rank \ac{TC} method, in which a discrete-space regularizer is added to the conventional \ac{NN} minimization-based low-rank \ac{TC} paradigm that enforces discreteness in the objective of the problem.

To that extent, the discrete-space regularizer is relaxed by first replacing it with an arbitrarily-tight continuous and differentiable, but not convex, function, which is then convexized via \ac{FP}.
The method also makes use of a \ac{PG} algorithm designed specifically for the formulated problem.

The contributions of this article can be summarized as follows:

\begin{itemize}
    \item A novel discrete-aware low-rank tensor completion method is proposed, where the discrete regularizer is based on an $\ell_0$-norm approximation, convexized via \ac{FP}, and solved via a \ac{PG} algorithm.
    \item A comprehensive performance comparison with \ac{SotA} methods is provided, demonstrating the superior performance of the proposed method in terms of \ac{NMSE} and convergence speed.
    %\item {\color{red}A detailed complexity analysis of the proposed method is presented, highlighting its computational efficiency compared to existing techniques.}
\end{itemize}

The structure of the article is as follows.
First, basics on the \ac{TC} problem, as a generalization of the \ac{MC} problem, with a description of conventional \ac{TC} techniques, as well as a review of the discrete-aware variation of the \ac{MC} proposed in \cite{Iimori_2020} and \cite{Nic_Asilo_2024} is given in Section \ref{sec:prior}.
Then, the new method based on extending the \ac{NN} minimization-based by a discrete-aware regularizer via
the $\ell_0$-norm approximation is described in Section \ref{sec:ProposedTC}, followed by simulations results comparing our contribution with \ac{SotA} methods and a few concluding remarks, in Sections \ref{sec:results} and \ref{sec:conclusions}, respectively.

\section{Preliminaries}
\label{sec:prior}

Tensors are a multi-dimensional generalization of vectors and matrices, such that an N-mode tensor $\mathcal{T}$ can be written as
\begin{equation}
    \mathcal{T} \in \mathbb{R}^{I_1 \times I_2 \times \cdots \times I_N},
\end{equation}
where $I_n$ denotes the size of the tensor in mode $n$, such that a vector is a 1-mode tensor, a matrix is a 2-mode tensor, and so on.

\begin{figure}[h]
    \centering
    \includegraphics[width=1\columnwidth]{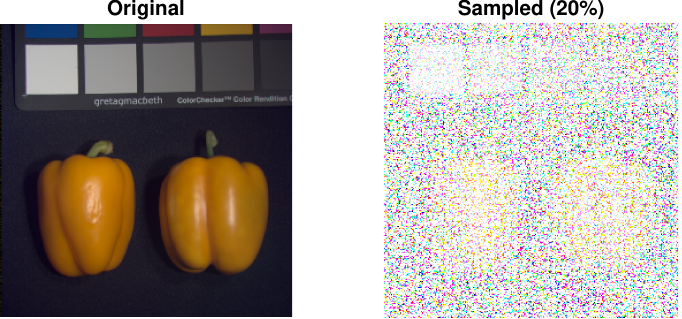}
    \vspace{-4ex}
    \caption{Illustration of the original tensor in form of an image, as well as its partially observed, sampled version, with $20\%$ observed entries.}
    \label{fig:og_image}
\end{figure}

For the sake of simplicity, we perform the tensor completion as part of image processing, as shown in Figure \ref{fig:og_image}, where a partially observed, sampled version of the image is received and needs to be recovered.
In this case, the image can be represented as a 3-mode tensor $\mathcal{T}\in \mathbb{R}^{W\times H\times 3}$, where the tensor is of size $W\times H\times3$, with $W$ and $H$ denoting the width and height of the image respectively and the third mode representing the RGB color channels.

\subsection{Classical Formulation of the Matrix and Tensor Completion Problem}
Before looking into the proposed method, a few major \ac{MC} techniques will be discussed that can be extended to tensor completion, all of which aim to recover entries from a partially observed N-mode tensor.
Typical \ac{SotA} low rank \ac{MC} techniques are based on a minimizing the rank of the matrix, such that the original rank minimization problem can be written as
\vspace{-1ex}
\begin{subequations}
\label{eq:rank_opt}
\begin{align}
\argmin_{\boldsymbol{X}\in \mathbb{R}^{m\times n}}&\quad \text{rank}(\boldsymbol{X}),\\
\text{s.t. }&\quad P_{\Omega}(\boldsymbol{X})=P_{\Omega}(\boldsymbol{O}),
\vspace{-1ex}
\end{align}
\end{subequations}
where $\text{rank}(\cdot)$ denotes the rank of the input matrix, and $P_{\Omega}(\cdot)$ indicates a mask operator, which is defined as 
\vspace{-1ex}
\begin{equation}
[P_{\Omega}(\boldsymbol{X})]_{i,j}=
\begin{cases}
[\boldsymbol{X}]_{i,j},& \text{if } (i,j) \in \Omega, \\
0,              & \text{otherwise},
\end{cases}
\vspace{-1ex}
\end{equation}
where $[\cdot]_{i,j}$ denotes the $(i,j)$-th element of a given matrix, and $\Omega$ defines the index set of that contains the indices of the observed elements.

Although problem \eqref{eq:rank_opt} is optimal, it has been shown that the problem is \ac{NP}-hard due to the non-convexity of the rank operator, such that its solution cannot be obtained efficiently, which applies to both, matrices and tensors \cite{Hillar_2013}.

Nevertheless, it has also been shown in \cite{Candes_2012}, however, that the rank minimization problem can be relaxed by replacing the rank objective with the \ac{NN} operator $||\boldsymbol{X}||_*$, which is given by the sum of the singular values of $\boldsymbol{X}$, since a matrix of rank $r$ has exactly $r$ nonzero singular values.
The relaxed optimization problem utilizing the \ac{NN} objective can be therefore written as
\begin{subequations}
\label{eq:NNproblem}
\begin{align}
\argmin_{\boldsymbol{X}\in \mathbb{R}^{m\times n}}&\quad ||\boldsymbol{X}||_*,\\
\text{s.t. }&\quad P_{\Omega}(\boldsymbol{X})=P_{\Omega}(\boldsymbol{O}) ,
\label{eq:C1}
\end{align}
\end{subequations}
where the \ac{NN} can be viewed as a tight lower bound of the rank operator \cite{Recht_2010}, compared to the original problem in \eqref{eq:rank_opt}.

%------------------------------------------------

In practical scenarios, the observed matrix typically contains measurement noise, in which case enforcing constraint \eqref{eq:C1} exactly is overly restrictive.
As a result, the low-rank requirement is commonly relaxed, and the target matrix is recovered as an approximately low-rank solution.
Thus, several related works, $e.g.$, \cite{OptSpace, RankEDM, Wong_2017}, have considered the following relaxed formulation of problem \eqref{eq:NNproblem}, defined as
\begin{subequations}
\label{eq:NN_opt}
\begin{align}
\argmin_{\boldsymbol{X}\in \mathbb{R}^{m\times n}}&\quad ||\boldsymbol{X}||_*,\\[-1ex]
\text{s.t. }&\quad \underbrace{\frac{1}{2}||P_{\Omega}(\boldsymbol{X}-\boldsymbol{O}) ||^2_F}_{\triangleq f(\boldsymbol{X})}\leq \epsilon,
\end{align}
\end{subequations}
where $f(\boldsymbol{X})$ was implicitly defined for notational convenience and will be used throughout the rest of the article.

Next, problem \eqref{eq:NN_opt} can be rewritten in regularized form as
\begin{equation}
\label{eq:NN_opt_reg}
\argmin_{\boldsymbol{X}\in \mathbb{R}^{m\times n}}\quad f(\boldsymbol{X})+\lambda||\boldsymbol{X}||_*,
\end{equation}
or, including the rank constraint directly, as
\begin{subequations}
\label{eq:rank_opt2}
\begin{align}
\argmin_{\boldsymbol{X}\in \mathbb{R}^{m\times n}}&\quad f(\boldsymbol{X}),\\
\text{s.t. }&\quad \text{rank}({\boldsymbol{X}})\leq \nu.
\end{align}
\end{subequations}

While many \ac{SotA} approaches \cite{Meka_2009,Hu_2013,Tanner_2015} focus on the solution of convex relaxations of the low-rank \ac{MC} problem, recent years have shown substantial progress in non-convex optimization methods \cite{Wang_2021,Sun_2016,Li_2018}.
By direct parameterization of the low-rank structure, these methods more closely approximate the original non-convex formulations and often achieve superior empirical performance when compared to convex alternatives.
A comprehensive overview of non-convex optimization techniques for low-rank \ac{MC} can be found in \cite{Chi_2019}.

%------------------------------------------------
As mentioned before, the same concept of nuclear norm minimization can be applied to Tensors with $\bm{\mathcal{X}},\bm{\mathcal{T}}=\mathbb{R}^{\bm{I}_1,\bm{I}_1,\cdots \bm{I}_N}$ being N-mode Tensors, such that
\begin{subequations}
\begin{align}
    \argmin &\quad||\boldsymbol{\mathcal{X}}||_*,\\
    \text{s.t} &\quad P_{\Omega}(\boldsymbol{\mathcal{X}})=P_{\Omega}(\boldsymbol{\mathcal{T}}), 
\end{align}
\end{subequations}

where the nuclear norm for a Tensor is defined as %\cite{Liu_2013}
\begin{equation}
    ||\boldsymbol{\mathcal{X}}||_*=\sum^N_{i=1}\gamma_i||\boldsymbol{X}_{(i)}||_*
\end{equation}
with $\gamma_i\geq 0,\sum{\gamma_i}=1$ and $\bm{X}_{(i)}$ being the mode i matricization , which can be rewritten as
\begin{subequations}
\begin{align}
    \argmin \sum^N_{i=1}\gamma_i &\quad||\boldsymbol{X}_{(i)}||_*,\\
    \text{s.t} &\quad P_{\Omega}(\boldsymbol{\mathcal{X}})=P_{\Omega}(\boldsymbol{\mathcal{T}}).
\end{align}
\end{subequations}

In general, mode-\(i\) matrizitation/flattening rearranges the elements of a tensor into a \ac{2D} matrix.
For a tensor \(\boldsymbol{\mathcal{T}} \in \mathbb{R}^{I_1 \times I_2 \times \cdots \times I_N}\), the mode-\(i\) unfolding is a matrix, defined as
\begin{equation}
    \boldsymbol{T}_{(i)} \in \mathbb{R}^{I_i \times \prod_{j \neq i} I_j}.
\end{equation}

To achieve the flattening, first, permute dimensions to bring the \(i\)-th dimension to the front and then reshape into a matrix.

The problem can then be solved by multiple methods, such as the one described in \cite{Liu_2013}.

%------------------------------------------------

A different matrix completion model can be based on matrix factorization \cite{xu2013parallel}, such that the optimization problem can be formulated as
\begin{subequations}
    \begin{align}
        \argmin_{\bm{X},\bm{U},\bm{V}} &\quad\frac{1}{2}||\bm{X}-\bm{U}\bm{V}^\intercal||_F^2,\\
        \text{s.t } &\quad P_{\Omega}(\boldsymbol{{X}})=P_{\Omega}(\boldsymbol{{T}}), 
    \end{align}
\end{subequations}
where $\bm{U}\in \mathbb{R}^{m\times r}$ and $\bm{V}\in \mathbb{R}^{n\times r}$, such that $\boldsymbol{X}\in \mathbb{R}^{m\times n}$
     
Similarly, an equivalent tensor completion version can be formulated, given by
\begin{subequations}
    \label{eq:MF_problem}
    \begin{align}
   \argmin_{\bm{X},\bm{Y}, \bm{\mathcal{X}}} & \quad \sum^N_{i=1}\frac{\gamma_i}{2}||\bm{X}_i\bm{Y}_i-\bm{X}_{(i)}||_F^2,\\
    \text{s.t } & \quad P_{\Omega}(\boldsymbol{\mathcal{X}})=P_{\Omega}(\boldsymbol{\mathcal{T}}), 
\end{align}
\end{subequations}
where $\bm{\mathcal{X}}\in \mathbb{R}^{I_1 \times I_2 \times \cdots \times I_N}$ and $\bm{X}_i\in \mathbb{R}^{I_i \times r_i}$, and $\boldsymbol{Y}_i\in \mathbb{R}^{r_i \times \prod_{j \neq i} I_j}$, which can be solved by conventional least square methods.

Another recent, more advanced SotA method \cite{Gao2018} proposes to combine the nuclear norm with matrix factorization, which leads to the problem given by
\begin{subequations}
    \begin{align}
    \argmin_{ \bm{\mathcal{X}}} & \quad\sum_{i\in \mathcal{A}}\gamma_i||\boldsymbol{X}_i||_*+\sum_{j\in \mathcal{B}}\frac{\gamma_i}{2}||\bm{X}_i\bm{Y}_i-\bm{X}_{(i)}||_F^2,\\
    \text{s.t } & \quad P_{\Omega}(\boldsymbol{\mathcal{X}})=P_{\Omega}(\boldsymbol{\mathcal{T}}), 
\end{align}
\end{subequations}
where \cite{Gao2018} proposes two solutions, a non-smooth version that relaxes the problem and a smooth version that uses a smooth approximation of the objective.

Finally, the discrete-aware matrix completion scheme first proposed in \cite{Iimori_2020} and later extended in \cite{Nic_Asilo_2024} assumes that the missing entries of the matrix that need be recovered belong to a finite discrete alphabet set $\mathcal{A}\triangleq\{a_1,a_2,\cdots\}$.
To that extent, the standard regularized optimization problem discussed in equation \eqref{eq:NN_opt_reg} can be extended to
\begin{equation}
    \argmin_{\boldsymbol{X}\in \mathbb{R}^{m\times n}}\quad f(\boldsymbol{X})+\lambda g(\boldsymbol{X})+\zeta r(\boldsymbol{X}|p),
\end{equation}
where $f(\boldsymbol{X})$ denotes the data fidelity term, $g(\boldsymbol{X})$ denotes a low-rank regularizer, which is chosen to be the \acf{NN}, while the discrete-space regularizer $r(\boldsymbol{X}|p)$ that adds discreteness to the problem is defined as
\begin{equation}
    r(\boldsymbol{X}|p)\triangleq \sum_{k=1}^{|\mathcal{A}|}||\text{vec}_{\bar{\Omega}}(\boldsymbol{X})-a_k\boldsymbol{1}||_p,
\end{equation}
with $0\leq p$, $|\mathcal{A}|$ defines the cardinality of the set of alphabet symbols and $\text{vec}_{\bar{\Omega}}(\boldsymbol{X})$ represents a vectorization of the matrix $\boldsymbol{X}$, where the entries are chosen corresponding to the given index set $\bar{\Omega}$, being the complementary set to the previously presented set $\Omega$.

\subsection{Classical Solution to the Matrix and Tensor Completion Problem}
Next, before the proposed method is presented, a few major \ac{MC} and \ac{TC} techniques will be discussed that can be extended to tensor completion, all of which aim to recover entries from a partially observed N-mode tensor.

\subsubsection{\Acl{SI} Approach}
\Ac{SI} \cite{Mazumder_2010} and its extension the accelerated and inexact \ac{SI} (AIS)-impute approach of \cite{Yao_2019}, are recently proposed methods to solve large-scale \ac{MC} problems \cite{Recht_2013,Fang_2017} that have a similar structure as the problems described by equations \eqref{eq:NN_opt} and \eqref{eq:NN_opt_reg}, which aim to solve the \ac{NN} minimization problem efficiently by using \ac{SVT}. 
The conventional \ac{SI} approach consists of the recursion
\begin{equation}
\label{eq:soft_inpute}
\boldsymbol{X}_t=\text{SVT}_\lambda(\boldsymbol{X}_{t-1}+P_{\Omega}(\boldsymbol{O}-\boldsymbol{X}_{t-1})),
\end{equation}
where the SVT function is defined as
\begin{equation}
\text{SVT}_\lambda(\boldsymbol{A})=\boldsymbol{U}(\boldsymbol{\Sigma}-\lambda\boldsymbol{I})_+\boldsymbol{V}^\intercal,
\end{equation}
with $(\cdot)_+$ being the positive part of the input and $\boldsymbol{A}\triangleq\boldsymbol{U}\boldsymbol{\Sigma}\boldsymbol{V}^\intercal$ .

Furthermore, it has been shown in \cite{Yao_2019} that \ac{SI} can be used as a \ac{PG} method, enabling the use of \ac{SI}-based Nesterov-type momentum acceleration \cite{Nesterov}, which is known to significantly speed up convergence.
The corresponding updated recursion can be written as
\begin{equation}
\boldsymbol{X}_t=\text{SVT}_\lambda(\boldsymbol{Y}_t+P_{\Omega}(\boldsymbol{O}-\boldsymbol{Y}_t)),
\end{equation}
where $\boldsymbol{Y}_t=(1+\gamma_t)\boldsymbol{X}_{t-1}+\gamma_t\boldsymbol{X}_{t-2}$ is moment acceleration function, and $\gamma_t$ denotes the weight of the momentum.

\subsubsection{Discrete-Aware Matrix Completion}

Following the two \ac{SotA} methods for discrete-aware matrix completion, a \ac{PG} method was used in both \cite{Iimori_2020} and \cite{Nic_Asilo_2024} to solve the problem iteratively, which results in the following steps recursion
\begin{eqnarray}
&\boldsymbol{Y}_t=(1+\gamma_t)\boldsymbol{X}_{t-1}+\gamma_t\boldsymbol{X}_{t-2},\label{eq:acc_old}&\\
&\boldsymbol{Z}_t=\text{prox}_{\zeta_r}(\boldsymbol{Y}_t),&\\
&\boldsymbol{X}_t=\text{SVT}_\lambda(P_{\bar{\Omega}}(\boldsymbol{Z}_t)+P_{\Omega}(\boldsymbol{O})),&
\end{eqnarray}
where the first step integrates the moment acceleration function as discussed in the (AIS)-impute approach, the second step corresponds to the proximal operation on the discrete space regularizer, which depends on the parameter $p$ used in the regularizer and the last step amounts to the \ac{PG} operation on the \ac{NN} regularizer, since, as discussed above, it was shown in \cite{Cai_2010} that \ac{SVT} can be used as a proximal minimizer for the \ac{NN}.

%------------------------------------------------
\subsubsection{Simple Low-Rank Tensor Completion (SiLRTC) Algorithm}
%{\color{red}Rewrite and change to FA}

The SiLRTC algorithm \cite{Liu_2013} addresses tensor completion through a convex relaxation framework. The basic formulation can be expressed as follows

\begin{subequations}
\begin{align}
\min _{\boldsymbol{\mathcal{X}}, \boldsymbol{M}_i} & \quad \sum_{i=1}^n \gamma_i\left\|\boldsymbol{M}_i\right\|_*, \\
\text { s.t. } & \quad \boldsymbol{\mathcal{X}}_{(i)}=\boldsymbol{M}_i, \quad i=1, \ldots, n, \\
& \quad P_{\Omega}(\boldsymbol{\mathcal{X}})=P_{\Omega}(\boldsymbol{\mathcal{T}}),
\end{align}
\end{subequations}
where $\gamma_i$ are the weights for each mode, $\boldsymbol{\mathcal{X}}_{(i)}$ denotes the mode-$i$ unfolding of the tensor $\boldsymbol{\mathcal{X}}$, and $\boldsymbol{M}_i$ are auxiliary variables introduced to facilitate the optimization.
To improve robustness and computational tractability, a relaxed version of the problem introduces a tolerance parameter $d_i$ for each mode, such that
\begin{subequations}
\begin{align}
\min _{\boldsymbol{\mathcal{X}}, \boldsymbol{M}_i} & \quad \sum_{i=1}^n \gamma_i\left\|\boldsymbol{M}_i\right\|_*, \\
\text { s.t. } & \quad \left\|\boldsymbol{\mathcal{X}}_{(i)}-\boldsymbol{M}_i\right\|_F^2 \leq d_i, \quad i=1, \ldots, n, \\
& \quad P_{\Omega}(\boldsymbol{\mathcal{X}})=P_{\Omega}(\boldsymbol{\mathcal{T}}).
\end{align}
\end{subequations}

An equivalent formulation can be obtained by introducing the penalty parameters $\beta_i$ associated with the quadratic constraints, leading to an unconstrained form, given by
\begin{subequations}
\begin{align}
\min _{\boldsymbol{\mathcal{X}}, \boldsymbol{M}_i} & \quad \sum_{i=1}^n \gamma_i\left\|\boldsymbol{M}_i\right\|_* + \frac{\beta_i}{2}\left\|\boldsymbol{\mathcal{X}}_{(i)}-\boldsymbol{M}_i\right\|_F^2, \\
\text { s.t. } & \quad P_{\Omega}(\boldsymbol{\mathcal{X}})=P_{\Omega}(\boldsymbol{\mathcal{T}}).
\end{align}
\end{subequations}

The minimization can be decomposed into two subproblems, such that for a fixed $\boldsymbol{M}_i$, the subproblem with respect to $\boldsymbol{\mathcal{X}}$ is given by
\begin{subequations}
\begin{align}
\min _{\boldsymbol{\mathcal{X}}} & \quad \sum_{i=1}^n \frac{\beta_i}{2}\left\|\boldsymbol{M}_i-\boldsymbol{\mathcal{X}}_{(i)}\right\|_F^2, \\
\text { s.t. } & \quad P_{\Omega}(\boldsymbol{\mathcal{X}})=P_{\Omega}(\boldsymbol{\mathcal{T}}),
\end{align}
\end{subequations}
which yields a closed-form solution, expressed as
\begin{equation}
\boldsymbol{\mathcal{X}}_{i_1, \ldots, i_n}= 
\begin{cases}
\left(\frac{\sum_i \beta_i \text{Fold}_i(\boldsymbol{M}_i)}{\sum_i \beta_i}\right)_{i_1, \ldots, i_n}, & (i_1, \ldots, i_n) \notin \Omega, \\
\boldsymbol{\mathcal{T}}_{i_1, \ldots, i_n}, & (i_1, \ldots, i_n) \in \Omega.
\end{cases}
\end{equation}

Next, each auxiliary variable $\boldsymbol{M}_i$ is updated by singular value thresholding

\begin{equation}
\boldsymbol{M}_i=\text{SVT}_{\frac{\gamma_i}{\beta_i}}\left(\boldsymbol{\mathcal{X}}_{(i)}\right),
\end{equation}
where the singular value thresholding operation is defined as in the previous subsection.

\subsubsection{Matrix Factorization Approach}

The objective function described in \eqref{eq:MF_problem} is convex with respect to each block of variables $\bm{X}$, $\bm{Y}$, and $\bm{\mathcal{X}}$ when the remaining two blocks are fixed.
Hence, a block coordinate descent scheme can be applied, as proposed in \cite{xu2013parallel}, where $\bm{X}$, $\bm{Y}$, and $\bm{\mathcal{X}}$ are iteratively updated.
At iteration $k$, the individual problems for the corresponding updates are given by
\begin{subequations}\label{eq:update}
\begin{align}
\bm{X}^{k+1}
&= \quad
\argmin_{\bm{X}} 
f(\bm{X},\bm{Y}^k,\bm{\mathcal{X}}^k),
\label{update-x}
\\
\bm{Y}^{k+1}
&= \quad
\argmin_{\bm{Y}}
f(\bm{X}^{k+1},\bm{Y},\bm{\mathcal{X}}^k),
\label{update-y}
\\
\bm{\mathcal{X}}^{k+1}
&= \quad\hspace{-2.5ex}
\argmin_{P_{\Omega}(\bm{\mathcal{X}})=P_{\Omega}(\bm{\mathcal{T}})}
f(\bm{X}^{k+1},\bm{Y}^{k+1},\bm{\mathcal{X}}).
\label{update-z}
\end{align}
\end{subequations}

Both subproblems \eqref{update-x} and \eqref{update-y} decompose into $N$ independent least-squares problems and admit parallel solutions.
The resulting closed-form updates can be expressed as
\begin{subequations}\label{eq:exupdate}
\begin{align}
\bm{X}_i^{k+1}
&=
\bm{X}_{(i)}^k (\bm{Y}_i^k)^\top
\big(
\bm{Y}_i^k (\bm{Y}_i^k)^\top
\big)^\dagger,
 i=1,\ldots,N,
\label{exupdate-x}
\\
\bm{Y}_i^{k+1}
&=\!
\big(
(\bm{X}_i^{k+1})^\top \bm{X}_i^{k+1}
\big)^\dagger
(\bm{X}_i^{k+1})^\top
\bm{X}_{(i)}^k,
 i=1,\ldots,N,
\label{exupdate-y}
\\
\bm{\mathcal{X}}^{k+1}
&=
P_{\Omega^c}
\left(
\sum_{i=1}^N
\text{Fold}_i
(
\bm{X}_i^{k+1}\bm{Y}_i^{k+1}
)
\right)
+
P_{\Omega}(\bm{\mathcal{T}}),
\label{exupdate-z}
\end{align}
\end{subequations}
where $(\cdot)^\dagger$ denotes the Moore--Penrose pseudo-inverse.

Since the recovery of $\bm{\mathcal{X}}$ depends exclusively on the products $\bm{X}_i \bm{Y}_i$, $i=1,\ldots,N$, the update for $\bm{X}_i$ can be simplified by omitting the pseudo-inverse in \eqref{exupdate-x}, leading to
\begin{equation}\label{exupdate2-x}
\bm{X}_i^{k+1}
=
\bm{X}_{(i)}^k (\bm{Y}_i^k)^\top,
\quad i=1,\ldots,N.
\end{equation}

Together with \eqref{exupdate-y}, this update yields identical products $\bm{X}_i^{k+1}\bm{Y}_i^{k+1}$ for all $i$ when compared to
\eqref{exupdate-x} and \eqref{exupdate-y}, which is proven in \cite[Lemma 2.1]{Wen_2012}.

%------------------------------------------------

% --------------------
% SECTION: PROPOSED METHOD
% --------------------
\section{Proposed Method}
\label{sec:ProposedTC}

In what follows, the proposed Discrete-Aware Tensor Completion model seeks to recover a tensor $\bm{\mathcal{X}}$ by solving an optimization problem composed of three distinct functions \cite{Nic_Asilo_2024}
\begin{equation}
    \argmin_{\bm{\mathcal{X}}} \underbrace{f(\bm{\mathcal{X}})}_{\text{Data Fidelity}} + \underbrace{\zeta h(\bm{\mathcal{X}})}_{\text{Discrete Prior}} + \underbrace{\lambda g(\bm{\mathcal{X}})}_{\text{Low-Rank Prior}},
\end{equation}
subject to the observation constraint $P_{\Omega}(\bm{\mathcal{X}}) = P_{\Omega}(\bm{\mathcal{T}})$. In this formulation, $f(\bm{\mathcal{X}}) = \frac{1}{2}\|P_{\Omega}(\bm{\mathcal{X}}-\bm{\mathcal{O}})\|_F^2$ ensures fidelity to the observed data \cite{Nic_Asilo_2024}. The low-rank prior $g(\bm{\mathcal{X}}) = \sum_{i \in N} \gamma_i \| \bm{\mathcal{X}}_{(i)}\|_*$ represents the Sum of Nuclear Norms (SNN), promoting a multi-linear low-rank structure \cite{Nic_Asilo_2024}. The term $h(\bm{\mathcal{X}})$ is the discrete-aware regularizer later derived from an $l_0$-norm approximation \cite{Nic_Asilo_2024}.
While the discrete-aware regularizer is similar to the discrete-aware \ac{MC}, it can be rewritten for the tensor completion problem, given by 
    \begin{equation}
{r( \bm{\mathcal{X}}_{(i)}|{p})}\triangleq \sum_{k=1}^{|\mathbb{A}|}||\text{vec}_{\Bar{\Omega}}( \bm{\mathcal{X}}_{(i)})-a_k\boldsymbol{1}||_{{p}},
\end{equation}
which can be used since the mode-i flattening does not change the entries of the tensor.

\begin{figure}[h]
    \centering
    \includegraphics[width=1\columnwidth]{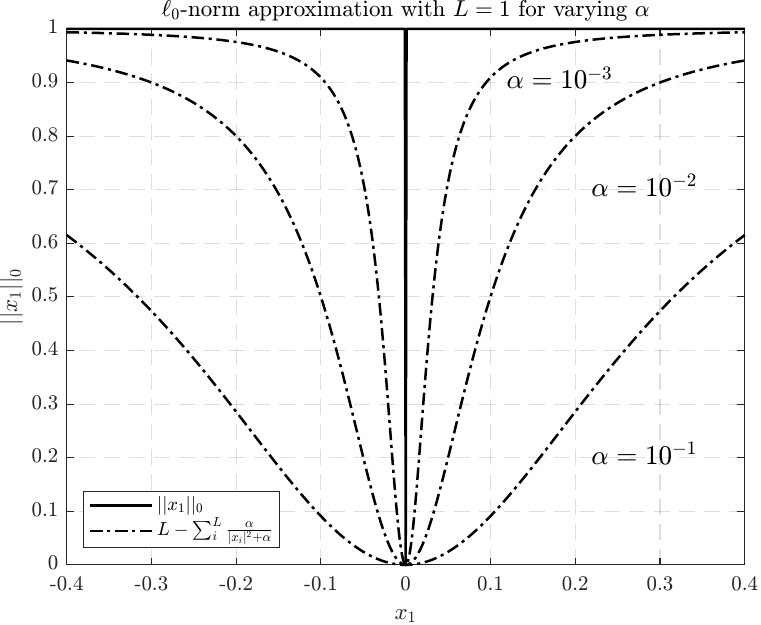}
    \vspace{-3ex}
    \caption{Illustration of the $l_0$-norm approximation with $L=1$ and a varying smoothing factor $\alpha$.}
    \label{fig:l0_approx}
\end{figure}

\subsection{Discrete Regularization via $l_0$-Norm Approximation}

Consider the following tight approximation of the $\ell_0$-norm proposed in \cite{Mohimani_2009} 
\begin{equation}
||\boldsymbol{x}||_0\approx\sum^L_{i=1}\frac{|x_i|^2}{|x_i|^2+\alpha}=L-\sum^L_{i=1}\frac{\alpha}{|x_i|^2+\alpha},
\end{equation}
where $L$ and $\alpha$ are the length of the vector $\boldsymbol{x}$ and an arbitrary small constant respectively.

Notice in particular, as illustrated in Figure~\ref{fig:l0_approx}, that under this approximation we have, for any value of $x_i$,
\vspace{-1ex}
\begin{eqnarray}
&\dfrac{|x_i|^2}{|x_i|^2+\alpha}\rightarrow 0 \text{ for } |x_i|^2=0,&\\
&\dfrac{|x_i|^2}{|x_i|^2+\alpha}\rightarrow 1 \text{ for } |x_i|^2>0.&
\end{eqnarray}

In order to apply the approximation to the discrete-aware regularizer, we first rewrite it in an element-by-element format, using the approximation to each term.
Proceeding as such, the regularizer can be written as
\vspace{-1ex}
\begin{align}
\label{eq:r_l0}
r(\boldsymbol{X}|0)&=\sum_{k=1}^{|\mathcal{A}|}||\text{vec}_{\bar{\Omega}}(\boldsymbol{X})-a_k\boldsymbol{1}||_0=\sum_{k=1}^{|\mathcal{A}|}\sum_{j=1}^{|\bar{\Omega}|}||\boldsymbol{x}_j-a_k||_0\nonumber \\ &\approx\sum_{k=1}^{|\mathcal{A}|}\bigl(\underbrace{|\bar{\Omega}|}_{const.}-\sum_{j=1}^{|\bar{\Omega}|}\frac{\alpha}{|\boldsymbol{x}_j-a_k|^2+\alpha}\bigr)\nonumber\\
&=-\sum_{k=1}^{|\mathcal{A}|}\sum_{j=1}^{|\bar{\Omega}|}\frac{\alpha}{|\boldsymbol{x}_j-a_k|^2+\alpha},
\end{align}
where $|\bar{\Omega}|$ denotes the number of elements in the set $\bar{\Omega}$.
Additionally, since $|\bar{\Omega}|$ is a constant term, it can be neglected in the following due to working on an optimization problem.

%------------------------------------------------
%\vspace{-2ex}
\subsection{Reformulation via Fractional Programming}

Notice that the discrete-aware regularizer formulation of equation \eqref{eq:r_l0} is not convex due to the fraction of affine and convex functions.
In order to mitigate this challenge, we introduce a fractional-programming variation of the $\ell_0$-norm approximation, enabled by the quadratic transform described in \cite{Shen_2018P1, Shen_2018P2}.

A simple reformulation can be found by applying the quadratic transform to the original $\ell_0$-norm approximation, which yields
\vspace{-1ex}
\begin{eqnarray}
\label{eq:l0_fp}
||x||_0 \hspace{-4ex}&& \approx\! L\!-\!\sum^L_{i=1}\frac{\alpha}{|x_i|^2\!+\!\alpha}\!\approx\!L\!-\!\bigg(\sum^L_{i=1}2\beta_i \sqrt{\alpha}\!-\!\beta_i^2(|x_i|^2\!+\!\alpha)\!\bigg)\nonumber\\[-1ex]
&&=\!\sum^L_{i=1}\!\beta_i^2|x_i|^2\!+\!\underbrace{L\!-\!\bigg(\!\sum^L_{i=1}2\beta_i \sqrt{\alpha}\!+\!\alpha\!\bigg)}_{\text{indep. of $x$}}\!\equiv\! \sum^L_{i=1}\beta_i^2|x_i|^2\!\!,\,
\end{eqnarray}
where $\beta_i\triangleq\frac{\sqrt{\alpha}}{|x_i|^2+\alpha}$ is an auxiliary variable resulting from the fractional programming step, and the terms independent of $x$ can be neglected for the purpose of the minimization problem, leading to the last equivalent.

In the following, the quadratic transform can be applied to the problem statement of the discrete-space regularizer similar as shown in \eqref{eq:l0_fp}, leading to
\begin{equation}
-\sum_{k=1}^{|\mathcal{A}|}\sum_{j=1}^{|\bar{\Omega}|}\frac{\alpha}{|\boldsymbol{x}_j-a_k|^2+\alpha}=\sum_{k=1}^{|\mathcal{A}|}\sum_{j=1}^{|\bar{\Omega}|}\beta_{k,j}^2|\boldsymbol{x}_j-a_k|^2, 
\end{equation}
with $\beta_{k,j}\triangleq\frac{\sqrt{\alpha}}{|\boldsymbol{x}_j-a_k|^2+\alpha}$ being an auxiliary variable resulting from the quadratic transform as stated above.

Consequently, the discrete-aware regularizer can be reformulated as
\begin{equation}
\label{eq:convexregularizer}
r(\bm{\mathcal{X}}|0)\!\approx\! h(\bm{\mathcal{X}})\!=\! \text{vec}_{\bar{\Omega}}(\bm{\mathcal{X}})^\intercal\! \boldsymbol{B}\text{vec}_{\bar{\Omega}}(\bm{\mathcal{X}})\! -\!2\text{vec}_{\bar{\Omega}}(\bm{\mathcal{X}})^\intercal \boldsymbol{b},
\end{equation}
with the auxiliary variables defined as
\begin{subequations}
\begin{equation}
    \bm{b} = \sum_{k=1}^{|\mathbb{A}|} a_k [\beta_{k,1}^2, \beta_{k,2}^2, \dots, \beta_{k,|\bar{\Omega}|}^2]^T \in \mathbb{R}^{|\bar{\Omega}|},
\end{equation}
\begin{equation}
    \bm{B} = \sum_{k=1}^{|\mathbb{A}|} \text{diag}[\beta_{k,1}^2, \beta_{k,2}^2, \dots, \beta_{k,|\bar{\Omega}|}^2]^T \in \mathbb{R}^{|\bar{\Omega}| \times |\bar{\Omega}|},
\end{equation}
\label{eq:updateBb}
\end{subequations}
where 
\begin{equation}
\beta_{k,j} = \frac{\sqrt{\alpha}}{|\text{vec}_{\bar{\Omega}}(\bm{\mathcal{X}}_j) - a_k|^2 + \alpha}.
\end{equation}

\subsection{Iterative Solution via Proximal Gradient}

Having obtained a closed-form and convex formulation of the approximate $\ell_0$-norm regularizer, the proposed optimization problem can be written as
\begin{equation}
\argmin_{\boldsymbol{\mathcal{X}}} {\underbrace{f(\boldsymbol{\mathcal{X}})}_{\mathclap{\frac{1}{2}||P_{\Omega}(\boldsymbol{\mathcal{X}}-\boldsymbol{O}) ||^2_F}}}+\lambda {\overbrace{g(\boldsymbol{\mathcal{X}})}^{\mathclap{||\boldsymbol{\mathcal{X}}||_*}}}+\zeta \underbrace{r(\boldsymbol{\mathcal{X}}|0)}_{\text{eq. \eqref{eq:convexregularizer}}},
\end{equation}
or, using equation \eqref{eq:convexregularizer}, as
\begin{equation}
\label{eq:finalproblem}
\argmin_{\boldsymbol{\mathcal{X}}} {f(\boldsymbol{\mathcal{X}})+{\zeta h(\boldsymbol{\mathcal{X}})}}+\lambda g(\boldsymbol{\mathcal{X}}).
\end{equation}

% Next, we process the SVT step where, as shown in \cite{antonello2020proximal}, the input corresponds to the matrix $\boldsymbol{X}_t$ and the gradient $\nabla h(\boldsymbol{X})$, which yields
% %
% \begin{align}
% \boldsymbol{X}_{t+1}&=\text{SVT}_\lambda(\boldsymbol{X}_{t}-\nabla f(\boldsymbol{X}_t)-\mu \zeta\nabla h(\boldsymbol{X}_t)) \nonumber \\
% &=\text{SVT}_\lambda(P_\Omega(\boldsymbol{O})+P_{\bar{\Omega}}(\boldsymbol{X}_{t}-\mu \zeta \nabla h(\boldsymbol{X_t}))),
% \label{eq:SVT_new}
% \end{align}
%

In order to solve the proposed optimization problem, we adapt the proximal gradient algorithm from \cite{Nic_Asilo_2024} to accommodate the discrete-aware regularizer and the tensor structure.

The proximal gradient method iteratively updates the estimate of the tensor $\bm{\mathcal{X}}$ by combining gradient descent steps for the smooth components and proximal operations for the non-smooth components \cite{Nic_Asilo_2024}, in addition to the discrete-aware regularizer.
At each iteration $t$, to enforce data fidelity, a gradient descent step is performed on the unobserved entries, resulting in the intermediate tensor
\begin{equation}
    \label{eq:gradient_step}
    \bm{{\mathcal{Y}}}_{t} = P_{\Omega}(\bm{\mathcal{O}}) + P_{\bar{\Omega}}({\bm{\mathcal{X}}}_t - \mu_t \zeta \nabla h({\bm{\mathcal{X}}}_t)),
\end{equation}
where $\mu$ denotes the a stepsize optimized for convergence guarantees by employing the Lipschitz condition
\begin{subequations}
\begin{equation}
0\leq\mu\leq 1/{S},
\label{eq:step}
\end{equation}
with $S$ denoting the Lipschitz constant, computed as
\label{eq:Lip}
\begin{equation}
S =\max\mathrm{eig} (\boldsymbol{B}^\intercal \boldsymbol{B}) =\mathrm{max}(B_{i,i}^2, \forall i ), \label{eq:Lip_low_comp}
\end{equation}
\end{subequations}
where $\max\mathrm{eig} (\cdot)$ denotes the operator returning the maximum eigenvalue of the argument matrix, and \eqref{eq:Lip_low_comp} offers a low complexity alternative.

In addition, the scalar $\zeta$ weights the influence of the discrete-aware regularizer in the gradient step, which is trivially computed from \eqref{eq:convexregularizer} as
\begin{equation}
    \label{eq:gradh}
    \nabla h(\bm{\mathcal{X}}) = 2 (\text{vec}_{\bar{\Omega}}(\bm{B}) \odot\text{vec}_{\bar{\Omega}}(\bm{\mathcal{X}}) - \bm{b}).
\end{equation}

With the intermediate tensor $\bm{\mathcal{Y}}_t$ computed, the low-rank prior is enforced via the proximal operator of the \ac{NN}, which can be efficiently computed on each mode unfolding of the tensor
\begin{equation}
    \label{eq:svt_step}
    \bm{\mathcal{X}}_{t+1} = \sum_{n=1}^N \alpha_n \text{Fold}_n \left( \text{SVT}_{\alpha_n \lambda} (\text{Unfold}_n(\bm{\mathcal{Y}}_t)) \right),
\end{equation}
where $\text{SVT}_{\tau}(\cdot)$ denotes the singular value thresholding operator with threshold $\tau$ \cite{Cai_2010}, which in our case is $\alpha_n \lambda$ for each mode $n$, where $\alpha_n$ are the weights for each mode satisfying $\sum_{n=1}^N \alpha_n = 1$ while $\lambda$ is another hyperparameter that specifies the weight of the low-rank prior.

To speed up convergence, we incorporate Nesterov-type momentum acceleration \cite{Nesterov} into the update process. The extrapolated point is computed as
\begin{equation}
    \label{eq:extrapolated}
    \hat{\bm{\mathcal{X}}}_t = (1 + \theta_t) \bm{\mathcal{X}}_{t-1} - \theta_t \bm{\mathcal{X}}_{t-2},
\end{equation}

where the momentum weight $\theta_t$ is updated with each iteration as $\theta_t = \frac{t-1}{t+2}$, as defined in \cite{Nesterov}.

The detailed procedure to solve problem \eqref{eq:finalproblem} is summarized as pseudo-code in Algorithm~\ref{alg:l0}.
The algorithm proceeds iteratively and terminates either after a maximum number of iterations or once the relative change between successive tensor estimates falls below a prescribed tolerance.
At initialization, the missing entries of the incomplete tensor are filled with the mean of the observed values, which yields the initial estimate $\bm{\mathcal{X}}_0$.
At each iteration, the algorithm first computes the auxiliary variables $\bm{b}$ and $\bm{B}$ for the unobserved entries based on the current tensor estimate. 
Using these quantities, an extrapolated tensor $\bm{\hat{\mathcal{X}}}_t$ is obtained via a Nesterov-type acceleration step. The gradient of the smooth objective function $h(\cdot)$ is then evaluated at $\bm{\hat{\mathcal{X}}}_t$, and the step size $\mu_t$ is set as the inverse of the maximum diagonal element of $\bm{B}$, according to the Lipschitz condition. 
A gradient descent step produces an intermediate tensor $\bm{\mathcal{Y}}_t$, after which a singular value thresholding operation updates the tensor estimate $\bm{\mathcal{X}}_{t+1}$.
The iteration stops once the relative Frobenius norm difference between two consecutive tensor estimates is below the tolerance $\epsilon$, and the final tensor estimate is returned as the output.

\begin{algorithm}[H]
\caption{Discrete-Aware Low-Rank Tensor Completion (DaLRTC)}
\begin{algorithmic}[1]
\small
\Statex \hspace{-4ex} \textbf{Input:} Incomplete tensor $\bm{\mathcal{O}}$, index set $\Omega$, 
auxiliary variable $\boldsymbol{\gamma}$, 
maximum iterations $T_{\max}$, tolerance $\epsilon$, discrete alphabet set $\mathbf{a}$, and weight $\lambda$.
\Statex \hspace{-4.4ex} \hrulefill \vspace{-0.3ex}
\State \textbf{Initialization:} 
Fill missing entries in $\bm{\mathcal{O}}$ with the mean of observed elements and set $\mathcal{X}_0=\mathcal{X}_{-1} = \bm{\mathcal{O}}$.

\For{$t = 1$ \textbf{to} $T_{\max}$}
    \State Compute auxiliary variables for unobserved entries $j \in \bar{\Omega}$:
    \State Compute $\bm{b}$ and $\bm{B}$ via equations \eqref{eq:updateBb}

    \State Update $\bm{\hat{\mathcal{X}}}_t$ via equation \eqref{eq:extrapolated} 

    \State Compute $\nabla h(\bm{\hat{\mathcal{X}}}_t)$ via equation \eqref{eq:gradh}
    \State $\mu_t \leftarrow 1 / \max(\text{diag}(\mathbf{B}))$
    \State Compute $\bm{\mathcal{Y}}_t$ using equation \eqref{eq:gradient_step}

    \State Update $\bm{\mathcal{X}_{t+1}}$ via equation \eqref{eq:svt_step}

    \If{$\|\bm{\mathcal{X}}_{t+1} - \bm{\mathcal{X}}_{t}\|_F / \|\bm{\mathcal{X}}_{t}\|_F < \epsilon$}
        \State \textbf{break}
    \EndIf
\EndFor
\State \textbf{return} $\bm{\mathcal{X}}^*_t$
\end{algorithmic}
\label{alg:l0}
\end{algorithm}
%------------------------------------------------
\section{Numerical Results}
\label{sec:results}

In this section we compare the proposed method to the \ac{NN} minimization-based \ac{SotA} technique (FaLRTC) described in \cite{Liu_2013}, the matrix factorization approach (TMac) of \cite{xu2013parallel}, and a recent technique that uses a mixture of both objectives (S-LRTC), as described in \cite{Gao2018}.
For comparison, the image\footnote{The test image is available at:\\ https://github.com/shangqigao/TensorCompletion} presented in Figure \ref{fig:og_image} is chosen, which is of size $256\times 256$ and contains RGB color channels represented as a third dimension, leading to a tensor of size $256\times 256\times 3$ and a finite discrete alphabet set $\mathcal{A}=\{0,1,\ldots,255\}$.
The chosen performance metric for comparison is the \ac{NMSE}, defined by
\begin{equation}
\mathrm{NMSE} = \frac{\|\widehat{\mathcal{X}} - \mathcal{X}_{\mathrm{true}}\|_F^2}{\|\mathcal{X}_{\mathrm{true}}\|_F^2}.
\label{eq:NMSE}
\end{equation}

\begin{figure}[H]
    \centering
    \includegraphics[width=\columnwidth]{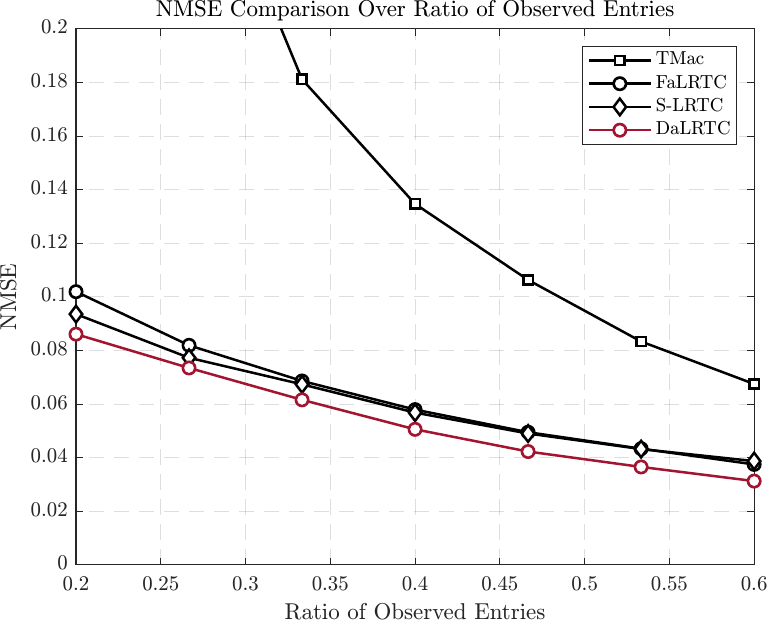}
    \vspace{-4ex}
    \caption{\ac{NMSE} comparison of the \ac{SotA} and the proposed method, with a varying ratio of observed entries in $\boldsymbol{O}$, for $\alpha=0.01$, $\lambda=65$ and $\zeta=0.5$.}
    \label{fig:NMSE}
\vspace{2ex}
    \includegraphics[width=\columnwidth]{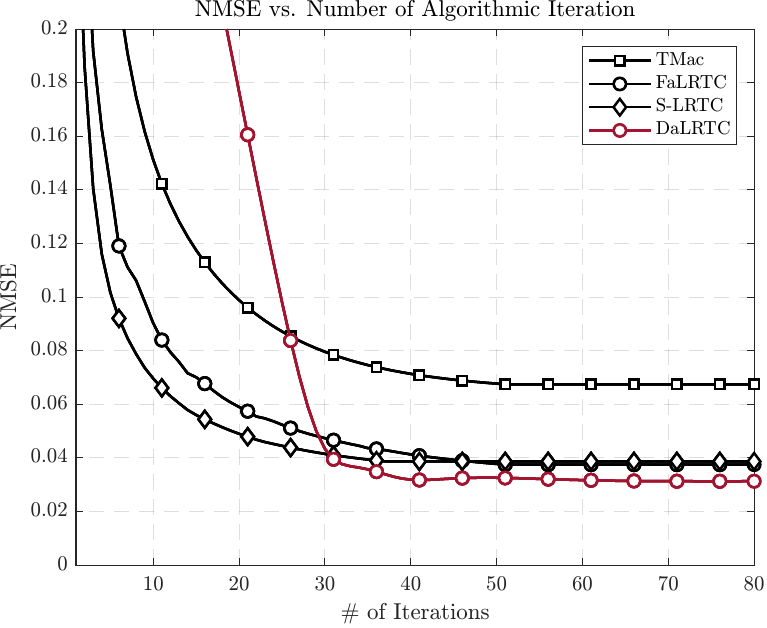}
    \vspace{-4ex}
    \caption{\ac{NMSE} convergence comparison of the \ac{SotA} and the proposed method, with a $60\%$ observation ratio of $\boldsymbol{O}$, for $\alpha=0.01$, $\lambda=65$ and $\zeta=0.5$.}
    \label{fig:conv}
\vspace{2ex}
    \includegraphics[width=\columnwidth]{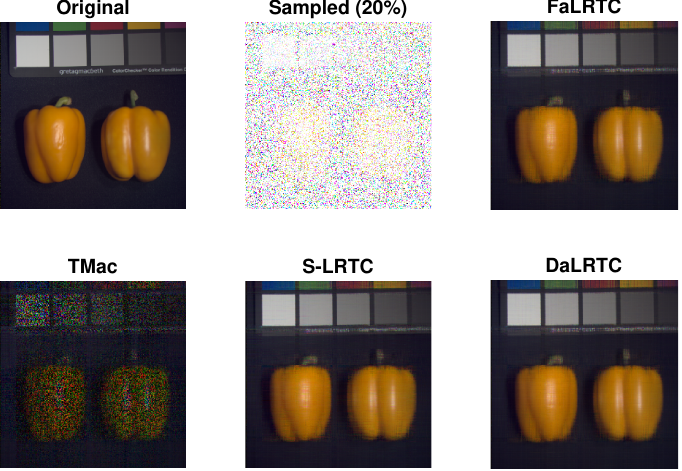}
    \vspace{-4ex}
    \caption{Illustration of the completed test image for the proposed method, compared to the original image, the sampled image and the \ac{SotA} methods reconstructed images for an observation ratio of $20\%$.}
    \label{fig:image}
\end{figure}

In order to compare all \ac{SotA} methods and the proposed method to the same conditions, identical simulation setups are chosen, where the ratio of observed values in the tensor $\boldsymbol{O}$ varies from $20\%$ to $60\%$.
The \ac{NMSE} results over varying observation ratios are shown in Figure \ref{fig:NMSE}, while the convergence behavior for a fixed observation ratio of $60\%$ is shown in Figure \ref{fig:conv}.
%
% Additionally, to further illustrate the convergence behavior of the proposed method, we also show in Figure \ref{fig:conv_extd} the convergence when initializing the proposed method with the solution of the earlier method from \cite{Gao2018}.

In Figure \ref{fig:NMSE}, it can be observed that the proposed method outperforms all \ac{SotA} methods.
While there is already a gain in adding discreteness to the nuclear norm objective, the proposed methods is also superior to more recent methods such as the S-LRTC approach.
Even though the gain between the two methods seems small it is significant comparing how much information is needed to achieve the same result.
Considering an expected NMSE of $0.4$, it can be seen that the proposed method requires a $48\%$ observed tensor, while the best \ac{SotA} requires at least a $60\%$ observed tensor.

Next, in Figure \ref{fig:conv}, the convergence of the methods is shown, illustrating that the proposed method converges as fast as the \ac{SotA} methods, however, with a slower start of convergence.
This can be explained by the additional auxiliary variables that result from the fractional programming reformulation, which need to be updated at each iteration and need some iterations to stabilize. 

Finally, for the sake of completeness, Figure \ref{fig:image} shows the visual result of the \ac{SotA} methods compared to the proposed method for a fixed observation ratio of $20\%$.
It can be observed that the proposed method is able to reconstruct the image with higher visual quality, with less artifacts and a more accurate color representation than the \ac{SotA} methods.

% \begin{figure}[h]
%     \centering
%     \includegraphics[width=\columnwidth]{Figures/plot_Conv.pdf}
%     \vspace{-4ex}
%     \caption{{\color{red}This can become the same kcikstarted with the SotA}}
%     \label{fig:conv}
% \end{figure}

\section{Conclusion}
\label{sec:conclusions}

We proposed a new discrete-aware \ac{TC} algorithm in which, in addition to the conventional \ac{NN} minimization-based objective, a discrete-space $\ell_0$-norm regularizer is introduced to the objective function, which is later approximated with asymptotically tight precision.
The resulting non-convex problem is then convexized via fractional programming and solved efficiently via a proximal gradient method, which enables the inclusion of the discrete-space awareness into the tensor completion problem.
Simulation results demonstrate that the proposed scheme outperforms both \ac{SotA} techniques, which enables more accurate reconstruction of low-rank tensors with discrete-valued entries.
The contributed scheme can not only be applied to image processing problems, but also problems such as signal processing, machine learning and data analysis, as well as scientific computing.
Future work will focus on the improvement the proposed method, both in terms of \ac{NMSE} performance and convergence.
%-----------------REFERENCES---------------------------

% \clearpage
\bibliographystyle{IEEEtran}
\bibliography{IEEEabrv,ref_paper.bib}

\begin{thebibliography}{10}
\providecommand{\url}[1]{#1}
\csname url@samestyle\endcsname
\providecommand{\newblock}{\relax}
\providecommand{\bibinfo}[2]{#2}
\providecommand{\BIBentrySTDinterwordspacing}{\spaceskip=0pt\relax}
\providecommand{\BIBentryALTinterwordstretchfactor}{4}
\providecommand{\BIBentryALTinterwordspacing}{\spaceskip=\fontdimen2\font plus
\BIBentryALTinterwordstretchfactor\fontdimen3\font minus
  \fontdimen4\font\relax}
\providecommand{\BIBforeignlanguage}[2]{{%
\expandafter\ifx\csname l@#1\endcsname\relax
\typeout{** WARNING: IEEEtran.bst: No hyphenation pattern has been}%
\typeout{** loaded for the language `#1'. Using the pattern for}%
\typeout{** the default language instead.}%
\else
\language=\csname l@#1\endcsname
\fi
#2}}
\providecommand{\BIBdecl}{\relax}
\BIBdecl

\bibitem{Chen_2022}
Z.~Chen and S.~Wang, ``A review on matrix completion for recommender systems,''
  \emph{Knowledge and Information Systems}, vol.~64, no.~1, pp. 1--34, 2022.

\bibitem{Nguyen_2019_Loc}
L.~T. Nguyen, J.~Kim, S.~Kim, and B.~Shim, ``Localization of iot networks via
  low-rank matrix completion,'' \emph{IEEE Transactions on Communications},
  vol.~67, no.~8, pp. 5833--5847, 2019.

\bibitem{Vlachos_2018}
E.~Vlachos, G.~C. Alexandropoulos, and J.~Thompson, ``Massive {MIMO} channel
  estimation for millimeter wave systems via matrix completion,'' \emph{IEEE
  Signal Processing Letters}, vol.~25, no.~11, pp. 1675--1679, 2018.

\bibitem{Nguyen_2019}
L.~T. Nguyen, J.~Kim, and B.~Shim, ``Low-rank matrix completion: A contemporary
  survey,'' \emph{IEEE Access}, vol.~7, pp. 94\,215--94\,237, 2019.

\bibitem{Dai_2012}
W.~Dai, E.~Kerman, and O.~Milenkovic, ``A geometric approach to low-rank matrix
  completion,'' \emph{IEEE Transactions on Information Theory}, vol.~58, no.~1,
  pp. 237--247, 2012.

\bibitem{Bart_2013}
B.~Vandereycken, ``Low-rank matrix completion by riemannian optimization,''
  \emph{SIAM Journal on Optimization}, vol.~23, no.~2, pp. 1214--1236, 2013.

\bibitem{Candes_2009_Noise}
E.~J. Cand\`{e}s and Y.~Plan, ``Matrix completion with noise,''
  \emph{Proceedings of the IEEE}, vol.~98, no.~6, pp. 925--936, 2010.

\bibitem{Candes_2009}
E.~J. Cand\`{e}s and B.~Recht, ``Exact matrix completion via convex
  optimization,'' \emph{Foundations of Computational Mathematics}, vol.~9,
  no.~6, pp. 717--772, 2009.

\bibitem{Candes_2010}
E.~J. Cand\`{e}s and T.~Tao, ``The power of convex relaxation: Near-optimal
  matrix completion,'' \emph{IEEE Transactions on Information Theory}, vol.~56,
  no.~5, pp. 2053--2080, 2010.

\bibitem{Cai_2010}
J.-F. Cai, E.~J. Cand\`{e}s, and Z.~Shen, ``A singular value thresholding
  algorithm for matrix completion,'' \emph{SIAM Journal on Optimization},
  vol.~20, no.~4, pp. 1956--1982, 2010.

\bibitem{Iimori_2020}
H.~Iimori, G.~T.~F. de~Abreu, O.~Taghizadeh, and K.~Ishibashi, ``Discrete-aware
  matrix completion via proximal gradient,'' in \emph{2020 54th Asilomar
  Conference on Signals, Systems, and Computers}, 2020, pp. 1327--1332.

\bibitem{Nic_Asilo_2024}
N.~F{\"u}hrling, K.~Ando, G.~T. Freitas De~Abreu, D.~{González G.}, and
  O.~Gonsa, ``Discrete aware matrix completion via convexized $\ell_{0}$-norm
  approximation,'' in \emph{2024 58th Asilomar Conference on Signals, Systems,
  and Computers}, 2024, pp. 916--920.

\bibitem{Song_2019}
\BIBentryALTinterwordspacing
Q.~Song, H.~Ge, J.~Caverlee, and X.~Hu, ``Tensor completion algorithms in big
  data analytics,'' \emph{ACM Trans. Knowl. Discov. Data}, vol.~13, no.~1, Jan.
  2019. [Online]. Available: \url{https://doi.org/10.1145/3278607}
\BIBentrySTDinterwordspacing

\bibitem{Ji_2019}
Y.~Ji, Q.~Wang, X.~Li, and J.~Liu, ``A survey on tensor techniques and
  applications in machine learning,'' \emph{IEEE Access}, vol.~7, pp.
  162\,950--162\,990, 2019.

\bibitem{Wang_2025}
X.-Y. Wang, X.-P. Li, N.~D. Sidiropoulos, and H.~C. So, ``Tensor completion
  network for visual data,'' \emph{IEEE Transactions on Signal Processing},
  vol.~73, pp. 386--400, 2025.

\bibitem{Liu_2013}
J.~Liu, P.~Musialski, P.~Wonka, and J.~Ye, ``Tensor completion for estimating
  missing values in visual data,'' \emph{IEEE Transactions on Pattern Analysis
  and Machine Intelligence}, vol.~35, no.~1, pp. 208--220, 2013.

\bibitem{xu2013parallel}
Y.~Xu, R.~Hao, W.~Yin, and Z.~Su, ``Parallel matrix factorization for low-rank
  tensor completion,'' \emph{arXiv preprint arXiv:1312.1254}, 2013.

\bibitem{Gao2018}
\BIBentryALTinterwordspacing
S.~Gao and Q.~Fan, ``A mixture of nuclear norm and matrix factorization for
  tensor completion,'' \emph{Journal of Scientific Computing}, vol.~75, no.~1,
  pp. 43--64, 2018. [Online]. Available:
  \url{https://doi.org/10.1007/s10915-017-0521-9}
\BIBentrySTDinterwordspacing

\bibitem{Hillar_2013}
C.~J. Hillar and L.-H. Lim, ``Most tensor problems are {NP-}hard,''
  \emph{Journal of the ACM (JACM)}, vol.~60, no.~6, pp. 1--39, 2013.

\bibitem{Candes_2012}
E.~Cand\`{e}s and B.~Recht, ``Exact matrix completion via convex
  optimization,'' \emph{Communications of the ACM}, vol.~55, no.~6, pp.
  111--119, 2012.

\bibitem{Recht_2010}
B.~Recht, M.~Fazel, and P.~A. Parrilo, ``Guaranteed minimum-rank solutions of
  linear matrix equations via nuclear norm minimization,'' \emph{SIAM Review},
  vol.~52, no.~3, pp. 471--501, 2010.

\bibitem{OptSpace}
R.~H. Keshavan, A.~Montanari, and S.~Oh, ``Matrix completion from a few
  entries,'' \emph{IEEE Transactions on Information Theory}, vol.~56, no.~6,
  pp. 2980--2998, 2010.

\bibitem{RankEDM}
W.~Glunt, T.~L. Hayden, S.~Hong, and J.~Wells, ``An alternating projection
  algorithm for computing the nearest euclidean distance matrix,'' \emph{SIAM
  Journal on Matrix Analysis and Applications}, vol.~11, no.~4, pp. 589--600,
  1990.

\bibitem{Wong_2017}
R.~K.~W. Wong and T.~C.~M. Lee, ``Matrix completion with noisy entries and
  outliers,'' \emph{Journal of Machine Learning Research}, vol.~18, no. 147,
  pp. 1--25, 2017.

\bibitem{Meka_2009}
R.~Meka, P.~Jain, and I.~Dhillon, ``Guaranteed rank minimization via singular
  value projection,'' \emph{NIPS}, 09 2009.

\bibitem{Hu_2013}
Y.~Hu, D.~Zhang, J.~Ye, X.~Li, and X.~He, ``Fast and accurate matrix completion
  via truncated nuclear norm regularization,'' \emph{IEEE Transactions on
  Pattern Analysis and Machine Intelligence}, vol.~35, no.~9, pp. 2117--2130,
  2013.

\bibitem{Tanner_2015}
J.~Tanner and K.~Wei, ``Low rank matrix completion by alternating steepest
  descent methods,'' \emph{Applied and Computational Harmonic Analysis},
  vol.~40, 08 2015.

\bibitem{Wang_2021}
Y.~Wang, Q.~Yao, and J.~Kwok, ``A scalable, adaptive and sound nonconvex
  regularizer for low-rank matrix learning,'' 04 2021, pp. 1798--1808.

\bibitem{Sun_2016}
R.~Sun and Z.-Q. Luo, ``Guaranteed matrix completion via non-convex
  factorization,'' \emph{IEEE Transactions on Information Theory}, vol.~62,
  no.~11, pp. 6535--6579, 2016.

\bibitem{Li_2018}
Q.~Li, Z.~Zhu, and G.~Tang, ``The non-convex geometry of low-rank matrix
  optimization,'' \emph{Information and Inference: A Journal of the IMA},
  vol.~8, 01 2018.

\bibitem{Chi_2019}
Y.~Chi, Y.~M. Lu, and Y.~Chen, ``Nonconvex optimization meets low-rank matrix
  factorization: An overview,'' \emph{IEEE Transactions on Signal Processing},
  vol.~67, no.~20, pp. 5239--5269, 2019.

\bibitem{Mazumder_2010}
R.~Mazumder, T.~Hastie, and R.~Tibshirani, ``Spectral regularization algorithms
  for learning large incomplete matrices,'' \emph{Journal of Machine Learning
  Research}, vol.~11, no.~80, pp. 2287--2322, 2010.

\bibitem{Yao_2019}
Q.~Yao and J.~T. Kwok, ``Accelerated and inexact soft-impute for large-scale
  matrix and tensor completion,'' \emph{IEEE Transactions on Knowledge and Data
  Engineering}, vol.~31, no.~9, pp. 1665--1679, 2019.

\bibitem{Recht_2013}
B.~Recht and C.~R\'{e}, ``Parallel stochastic gradient algorithms for
  large-scale matrix completion,'' \emph{Mathematical Programming Computation},
  vol.~5, no.~2, pp. 201--226, 2013.

\bibitem{Fang_2017}
H.~Fang, Z.~Zhang, Y.~Shao, and C.-J. Hsieh, ``Improved bounded matrix
  completion for large-scale recommender systems.'' in \emph{IJCAI}, 2017, pp.
  1654--1660.

\bibitem{Nesterov}
Y.~E. {Nesterov}, ``A method for solving the convex programming problem with
  convergence rate {O(1/$k^2$)},'' \emph{{Dokl. Akad. Nauk SSSR}}, vol. 269,
  pp. 543--547, 1983.

\bibitem{Wen_2012}
Z.~Wen, W.~Yin, and Y.~Zhang, ``Solving a low-rank factorization model for
  matrix completion by a nonlinear successive over-relaxation algorithm,''
  \emph{Mathematical Programming Computation}, vol.~4, no.~4, pp. 333--361,
  2012.

\bibitem{Mohimani_2009}
H.~Mohimani, M.~Babaie-Zadeh, and C.~Jutten, ``A fast approach for overcomplete
  sparse decomposition based on smoothed $\ell^{0}$ norm,'' \emph{IEEE
  Transactions on Signal Processing}, vol.~57, no.~1, pp. 289--301, 2009.

\bibitem{Shen_2018P1}
K.~Shen and W.~Yu, ``Fractional programming for communication systems---part
  {I}: Power control and beamforming,'' \emph{IEEE Transactions on Signal
  Processing}, 2018.

\bibitem{Shen_2018P2}
------, ``Fractional programming for communication systems---part {II}: Uplink
  scheduling via matching,'' \emph{IEEE Transactions on Signal Processing},
  vol.~66, no.~10, pp. 2631--2644, 2018.

\end{thebibliography}

\vspace{-2ex}\vfill\pagebreak

\end{document}